# Broadband Microwave Absorption by Logarithmic Spiral Metasurface

Shubo Wang[1,*], Bo Hou[2,3], and C. T. Chan[4,*]

[1]*Department of Physics, City University of Hong Kong, Hong Kong, China*

[2]*School of Physical Science and Technology & Collaborative Innovation Center of Suzhou Nano Science and Technology, Soochow University, Suzhou 215006, China*

[3]*Key Laboratory of Modern Optical Technologies of Ministry of Education & Key Lab of Advanced Optical Manufacturing Technologies of Jiangsu Province, Suzhou 215006, China*

[4]*Department of Physics, The Hong Kong University of Science and Technology, Hong Kong, China*

\* Correspondence should be addressed to:

Shubo Wang (shubwang@cityu.edu.hk) or C. T. Chan (phchan@ust.hk)

**Abstract**

We propose to use logarithmic spiral resonators for efficient absorption of microwaves. By combining their scale invariant geometries and Fabry-Perot-type resonances stemming from the fundamental TM mode, we realize a microwave metasurface with broadband absorption performance. The metasurface comprises logarithmic spiral resonators backed with a metallic surface and it can absorb > 95% of incident microwave energy within the frequency range of 6 GHz – 37 GHz. We discuss the physics underlying the broadband absorption and the crucial role of vortex energy flow. The study opens a new direction of electromagnetic wave absorption by employing the scale invariance of Maxwell equations.

## I.   INTRODUCTION

The absorption of electromagnetic waves has continuously drawn attention due to various applications such as stealth, reduction in radiation exposure, and solar energy related technologies. Since the advent of metamaterials [1–3], a large number of studies have been carried out to explore the possibility of absorbing electromagnetic waves by using metamaterials [4–23]. In particular, several designs of metamaterials for achieving broadband absorption have been proposed, such as the dispersion-engineered metamaterial [24, 25] and the pyramid metamaterial [26, 27]. While these designs are geometrically quite different, they all involve the assembling and fine tuning of multiple structural components in one unit cell in order to induce multiple electromagnetic resonances to broaden the absorption bandwidth. A simple design recipe of absorbers that can achieve broadband strong absorption of electromagnetic waves would be highly desirable. We will show that the scale invariance of Maxwell equations provides a possible solution to this problem. Following the recipe, we designed a metasurface consisting of logarithmic spiral resonators (LSRs) which can absorb > 95% of incident microwave within the frequency range of 6 GHz – 37 GHz. In contrast, a metasurface formed of Archimedean spiral resonators (ASRs), which has no scale invariant feature, only absorbs microwave strongly at particular frequencies. Note that several metamaterials consisting of spiral structures have been proposed for electromagnetic wave absorption [28–32], but their performance is narrow-banded.

Consider the source-free Maxwell equations in vacuum: $\nabla^2 \mathbf{E}(\mathbf{r},t) = (1/c^2)\partial^2 \mathbf{E}(\mathbf{r},t)/\partial t^2$, $\nabla^2 \mathbf{B}(\mathbf{r},t) = (1/c^2)\partial^2 \mathbf{B}(\mathbf{r},t)/\partial t^2$. These equations are invariant under the transformation $\mathbf{r} \to \chi\mathbf{r}$, $t \to \chi t$, where $\chi$ is an arbitrary scaling factor. The scale invariance indicates that the electromagnetic response of a system remains unchanged under the scaling of both geometric dimensions and wavelength. In materials with weak dispersion, this invariance can be approximately maintained. Under this condition, it is possible to design a metamaterial with broadband properties (e.g. absorption) deriving from the scale invariance of its unit cell geometry. In contrast, the broadband properties of conventional metamaterials usually derive from multiple structures with different geometric dimensions, which induce electromagnetic resonances at different frequencies. We will show that metamaterials with a scale invariant unit cell geometry can kill two birds with one stone: they can not only achieve broadband absorption of electromagnetic waves, but also maintain a small thickness as the result of a space coiling feature.

## II. RESULTS AND DISCUSSION
### A. Logarithmic spiral resonator

We consider the two-dimensional metamaterial unit shown in Fig. 1 (a). The geometry of the unit (left of Fig. 1(a)) is a logarithmic spiral defined by the function $r(\theta) = \alpha e^{\beta\theta}$, where $r, \theta$ are the polar coordinates and $\alpha, \beta$ are two arbitrary constants. An isotropic scaling of the spiral by a factor of $\chi$ leads to the transformation of $r \to \chi r = \chi\alpha e^{\beta\theta} = \alpha e^{\beta(\theta + \ln\chi/\beta)}$, which is identical to the rotation of $r(\theta)$ by an angle of $\ln\chi/\beta$. Therefore, an arbitrary scaling operation on the logarithmic spiral followed by a rotation recovers the original geometry, i.e., it is scale invariant. We consider a logarithmic spiral made of copper with conductivity $\sigma = 5.96 \times 10^6 \text{S/m}$, filled with a dielectric material with relative permittivity $\varepsilon_r = 1.4 + i\delta$ (grey region). The copper film has a thickness of 0.1 mm. The spiral can be viewed as transformed from a tapered parallel-plate (TPP) waveguide through space coiling. Under TM polarization, the TPP waveguide has a fundamental mode that has no cutoff frequency, which is similar to the normal parallel-plate waveguide. Figure 1(b) shows the absorption of this mode in the TPP waveguide. The absorption is defined as $1 - |S_{11}|^2 - |S_{21}|^2$, where $S_{11}$ and $S_{22}$ are the S-parameters. The length of the waveguide along $x$ direction is ~21 mm. The width of the left and the right openings is 0.01mm and 3.4 mm, respectively. We consider port excitation at the right opening. The variation of the waveguide's width induces impedance mismatch and hence Fabry-Perot-type resonances, which accounts for the peaks in the absorption spectrum in Fig. 1(b). We notice that when the dielectric loss $\delta$ is large, the absorption can approach 100% in a wideband. However, the TPP waveguide is bulky. In contrast, the logarithmic spiral is geometrically compact and its scale invariance should give rise to similar absorption property at different frequencies. Figure 1(c) shows the normalized absorption cross section (ACS) of the LSR with $\alpha = 0.0031$ mm, $\beta = 0.22$ and $\theta \in [0, 33]$, which is calculated by full-wave simulations using COMSOL [33]. We note that when $\delta$ is small, multiple peaks appear in the ACS spectrum due to the Fabry-Perot-type resonances. As $\delta$ is increased, the ACS grows and can beat the theoretical upper limit of single channel subwavelength resonance (i.e. $\lambda/2\pi$ [34], marked by the dashed line) as in the case of $\delta = 0.2$ at $f = 6.7$ GHz. Since large ACS must be accompanied by large scattering cross section (SCS), the LSR has a large extinction cross section as shown in Fig. 1(d) for the magnetic field amplitude at $f = 6.7$ GHz. When $\delta$ is large enough, we obtain a broadband large ACS due to the scale invariance of the LSR.

### B. Logarithmic spiral metasurface

We now employ the strong absorption of the LSR to build a metasurface (i.e. thin-layer metamaterial) microwave absorber. This is achieved by arranging the LSRs into a one-dimensional array with period $a = 7.83$ mm on a copper substrate, as shown in Fig. 2(a). The opening of the LSRs are facing to the direction of incoming waves. The overall thickness of the metasurface is $d = 5$ mm. The grey region denotes lossy dielectric material with relative permittivity $\varepsilon_r = 1.4 + i$. We calculated the absorption of a TM-polarized plane wave normally incident on the metasurface. The results are shown in Fig. 2(b) as the solid red line. We note that the metasurface can absorb > 95% of incident microwave energy within the frequency range of 7.7 GHz – 37 GHz. The lower bound here is determined by the lowest order resonance of the LSR. The upper bound is determined by the period of the metasurface. Above 37 GHz, the absorption reduces due to the effect of high-order diffractions. For comparison, we also calculated the absorption of a homogeneous slab with relative permittivity $\varepsilon_r = 1.4 + i$, which is backed with a copper substrate. The results are shown in Fig. 2(b) as the solid blue line. The lowest frequency with 95% absorption for the slab is 12.8 GHz which is much higher than that of the LSR metasurface. Besides, the slab has a lower absorption of ~ 90% at 20 GHz – 30 GHz. Therefore, the LSR metasurface provides higher absorption than the slab absorber especially at lower frequencies, which enables the realization of thin microwave absorbers. The high absorption of the LSR metasurface is corroborated by the reflected magnetic field shown in Fig. 2(c) for the case of $f$ = 10 GHz. The field is mainly concentrated in the center of the LSR due to the Fabry-Perot-type resonances and almost no field is reflected back. We then studied the dependence of the absorption on the incident angle of the microwave. Figure 2(d) shows the numerically calculated absorption as a function of frequency and incident angle. We note that high absorption can be achieved for incident angles lying within [-30, 30] degrees. As the angle increases, the bandwidth of high absorption reduces. The dependence of absorption on the incident angle is attributed to the special symmetry of the LSR. We also note that the maximum absorption bandwidth appears in the case of normal incidence. In the cases of oblique incidence at high frequencies, high-order diffraction contributes to the reflected wave and causes a lower absorption.

If we optimize the LSR metasurface by using graded-index lossy dielectric as the filling material, we can further broaden its absorption bandwidth. Figure 3(a) shows one unit cell of the optimized LSR metasurface which has a period of $a$ = 7.45 mm and a thickness of $d$ = 5 mm. We set $\alpha = 0.008$ mm, $\beta = 0.2$, and $\theta \in [0, 31.4]$. The grayscale indicates the absolute value of the permittivity. The first layer of the LSR has $\varepsilon_r = 1 + \theta(\varepsilon_c - 1)/(2\pi)$ with $\varepsilon_c = 4.2 + 2.6i$. The inner layers have $\varepsilon_r = \varepsilon_c$. The material outside the LSR is homogenous with $\varepsilon_r = 2 + 1.5i$. For comparison, we also considered a slab absorber with graded permittivity $\varepsilon_r = 1 + h(\varepsilon_c - 1)/d$ as shown in Fig. 3(b), where $d = 5$ mm and $h$ is the distance from the upper surface of the slab. We considered two cases with $\varepsilon_c = 4.2 + 2.6i$ and $\varepsilon_c = 1.4 + i$, respectively. The absorption spectra of the optimized structures are shown in Fig. 3(c). The LSR metasurface can achieve > 95% absorption within 6 GHz – 37 GHz, which is much better than the two slab absorbers. We note that the theoretical limit on the thickness of a non-magnetic absorber can be determined with the Rozanov relation: $d \geq 1/(2\pi^2) \left| \int_0^\infty \ln|R(\lambda)| d\lambda \right|$, where $R$ is the reflection coefficient [35]. We found that the LSR metasurface is only 1 mm thicker than this limit.

The metasurface absorber in Fig. 2(a) is invariant along $y$ direction and mainly absorbs TM-polarized microwaves. To achieve high absorption for both TM and TE polarizations, we then design a metasurface consisting of a two-dimensional array of LSRs on a copper surface. The unit cell is shown in Fig. 4(a), where two identical LSRs are orthogonally arranged so that they can absorb both TM- and TE-polarized microwaves. Note that the LSRs are finite along axis direction with a thickness of $c = 2.5$ mm. The unit cell has a dimension of $a \times b \times d = 10.2 \text{ mm} \times 7.2 \text{ mm} \times 5.4 \text{ mm}$. We numerically calculated the absorption of the metasurface and the results are shown in Fig. 4(b). We note that for both $y$ (i.e. TM) and $x$ (i.e. TE) polarizations the absorption is above 90% within the frequency range of 8 GHz – 38 GHz. Compared to the previous case in Fig. 2(b), the absorption is lower due to the truncation of the LSRs along the center axis direction, which leads to smaller quality factors of the resonances and hence a relatively weaker enhancement. Figure 4(c) and (d) show the dependence of the absorption on the incident angle for both the $y$ and $x$ polarizations, respectively. In both cases, high absorption can be achieved for incident angle within [-30, 30] degrees. The different performances of the metasurface under $y$ and $x$ polarizations are due to the asymmetric orientation of the two LSRs in one unit cell.

### C. Archimedean spiral metasurface

To further understand the critical role of scale invariance in the broadband absorption, we performed a comparative study where the metasurface is constructed with ASRs. The ASR has a geometry that is not scale invariant, as shown by the inset in Fig. 5(a), which is defined by the function $r(\theta) = r_0 + \rho\theta$ with $r_0$ and $\rho$ being constants. We set $r_0 = 0$ for simplicity. The ASR contains five layers with an overall diameter of $D$ and is filled with dielectric material ($\varepsilon_r = 2 + i\delta$). Similar to the LSR, the ASR also supports Fabry-Perot-type resonances, which have been used to achieve unusual effective permeability in metamaterials [1]. The resonances can significantly enhance the ACS of the ASR, as shown in Fig. 5(a) for the enhancement by the lowest order resonance. In contrast to the LSR, the large ACS of the ASR is narrowband. As the dielectric loss $\delta$ is increased, the ACS first grows and then decreases. The maximum absorption, i.e. the single channel theoretical upper limit $\lambda/2\pi$, is achieved at the normalized frequency ~ $f = 0.0246\ c/D$ in the case of $\delta = 0.004$, where $c$ is the speed of light in vacuum. Figure 5(b) shows the magnetic field amplitude at the resonance frequency, where the ASR (marked by an arrow) induces a large extinction cross section. We then consider a metasurface absorber consisting of one-dimensional array of ASR and a reflection metal surface, as shown by the inset in Fig. 5(c). The period of the metasurface is 18$D$. The filling dielectric material has a relative permittivity of $\varepsilon_r = 2 + 0.005i$. We calculated the absorption of the ASR metasurface and the results are shown in Fig. 5(c) by the red solid line. Notice that near-perfect absorption can be achieved at the resonance frequency of 0.0242$c/D$. Figure 5(d) shows the Poynting vectors within one unit cell at this frequency. We notice that the ASR behaves like an energy sink (i.e. singularity) that absorbs all the microwave in one unit cell. To compare the ASR metasurface with the LSR metasurface, we then set $\rho = 0.09$ mm so that the ASR has approximately the same volume as the LSR. We assume equal period and thickness for the two types of metasurfaces. Both homogeneous filling and graded-index filling are considered for comparison. Figure 5(e) shows the ASR metasurface with graded-index filling, where the outer layer has permittivity $\varepsilon_r = 1 + \theta(\varepsilon_c - 1)/(2\pi)$ with $\varepsilon_c = 4.2 + 2.6i$ and the inner layers have $\varepsilon_r = \varepsilon_c$. The dielectric outside the ASR has $\varepsilon_r = 2 + 1.5i$. Figure 5(f) shows the absorption spectra of the ASR metasurfaces, which can only reach 80% above 9.3 GHz and are not comparable to that of the LSR metasurfaces in Figs. 2 and 3. This confirms the essential role of scale invariance in achieving broadband high absorption.

We note that the performance of the LSR metasurface depends on the values of $\alpha$ and $\beta$. The constant $\alpha$ controls the size of the spiral and determines the working frequency of the metasurface. The constant $\beta$ controls how "compact" the spiral is and determines the mode density (i.e. number of resonances per frequency interval) of the LSR. When $\beta$ takes an appropriate value, the corresponding mode density results in broadband matching of the metasurface impedance with vacuum impedance and therefore significantly improves absorption. Some metamaterials constructed with fractal structures can also absorb electromagnetic waves within a broadband of frequencies [36–39]. The fractal structures are self-similar and are invariant under discrete scaling, i.e. $\chi$ takes discrete values. Self-similarity can broaden the absorption spectra of metamaterials by providing multiple resonances at discrete frequencies. But the spectral distribution of the resonances has poor tunability, leading to poor matching between the metamaterial impedance and vacuum impedance. Therefore, fractal metamaterials usually have non-smooth spectra with relatively low absorption. In contrast, the logarithmic spiral is invariant under continuous scaling transformation, i.e. $\chi$ can take arbitrary values. The scale invariance enables easy tuning of the metasurface impedance via parameter $\beta$ and the absorption spectrum is not only broadband but also smooth. There might be other geometries that are scale invariant, but the LSR has a unique property which contributes to the strong absorption of electromagnetic waves: it induces vortex flow of electromagnetic energy. The vortex channel enables efficient absorption of energy by guiding waves to travel a long way in a compact space volume. This scenario is similar to the dissipation of kinetic energy in turbulent flows in fluid dynamics, where the kinetic energy of a fluid undergoes strong damping due to the hierarchical structure of vortices: the energy is transferred from large vortices to small vortices and eventually converted to heat due to viscosity [40]. Here in the LSR metasurface, electromagnetic energy flows from the outer layer of the LSR (corresponding to large vortex) to the inner layers (corresponding to smaller vortices), and is eventually absorbed in the center due to material loss. The similarity between the two completely different physical systems indicates the critical role of vortex flow in the energy dissipation of classical systems.

## III. CONCLUSION

In conclusion, we provided a general recipe for designing broadband electromagnetic wave absorbers by combining scale invariant geometry and no-cutoff TM guided mode. Following the recipe, we designed a metasurface by using LSRs which can achieve high absorption of incident microwave within a broadband of frequencies. The critical role of scale invariance is verified through a comparative study with ASR metasurface. The results may be extended to high frequency regime for the absorption of visible light. The related physics may also be applied to the absorption of other classical waves such as sound.

## ACKNOWLEDGMENTS

This work was supported by the Research Grants Council of Hong Kong SAR (No. AoE/P-02/12 and No. CityU 21302018). S. W. was also supported by grants from City University of Hong Kong (No. 7200549 and No. 9610388). B. H. was supported by the grant from Natural Science Foundation of China (NSFC) (No. 11474212) and the Priority Academic Program Development (PAPD) of Jiangsu Higher Education Institutions. We thank Prof. Z. Q. Zhang for valuable comments and suggestions.

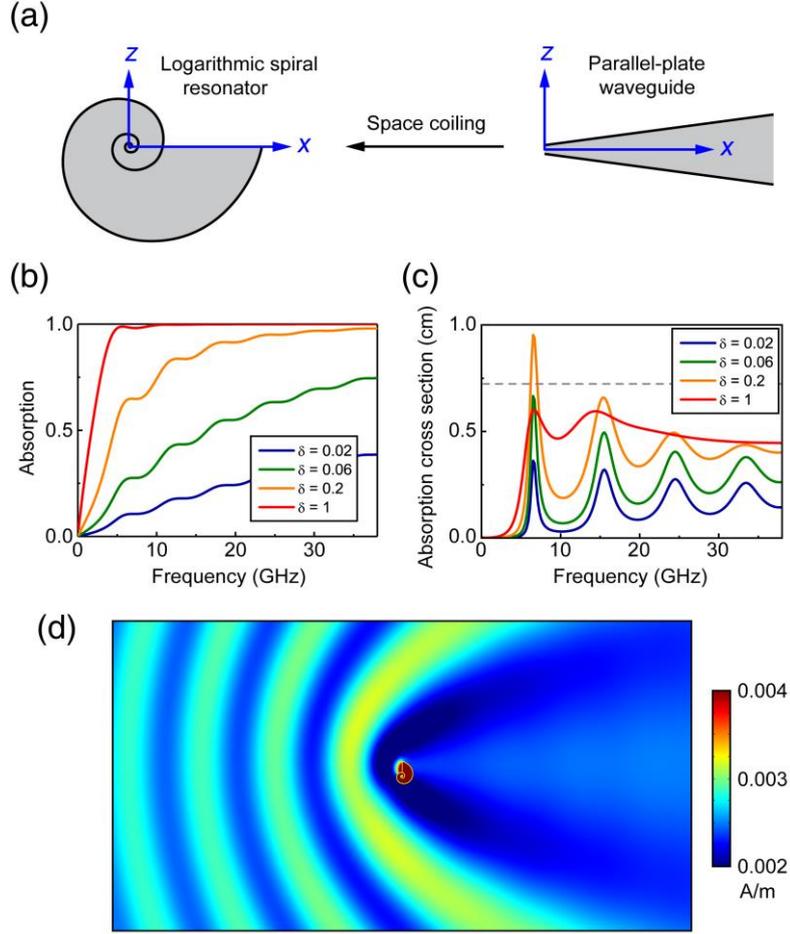

Fig. 1 (a) The proposed LSR (left) is drawn to scale. It can be viewed as the coiling of a tapered parallel-plate waveguide (right). We set $\alpha = 0.0031$ mm, $\beta = 0.22$, and $\theta \in [0, 33]$ (see text for the definition). (b) Absorption spectrum of the tapered parallel-plate waveguide for the fundamental TM mode under different loss of dielectric material ($\varepsilon_r = 1.4 + i\delta$). The width of the openings is 0.01 mm and 3.4 mm. The total length is 21 mm. (c) Normalized absorption cross section of the LSR for different loss parameter $\delta$. (d) Magnetic field amplitude at $f = 6.7$ GHz with $\delta = 0.2$.

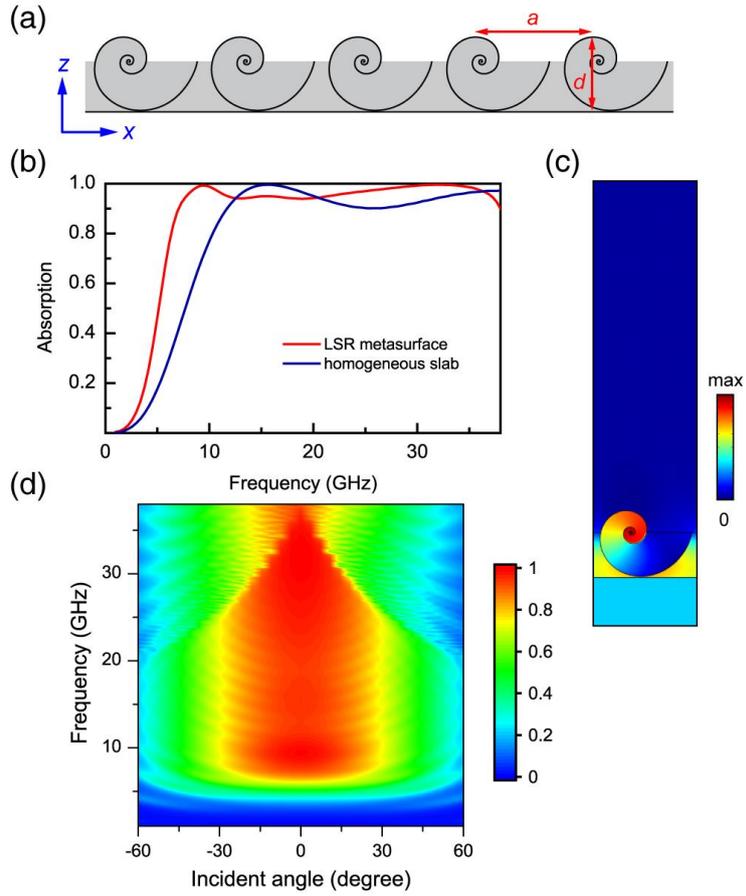

Fig. 2 (a) The LSR metasurface with thickness $d = 5$ mm and period $a = 7.83$ mm. The LSR is filled with dielectric material ($\varepsilon_r = 1.4 + i$). (b) Absorption of the LSR metasurface (red) and a homogeneous dielectric slab with $\varepsilon_r = 1.4 + i$ (blue) under normal incidence of a TM-polarized plane wave. (c) Reflected magnetic field amplitude at $f = 10$ GHz. (d) Absorption as a function of frequency and incident angle.

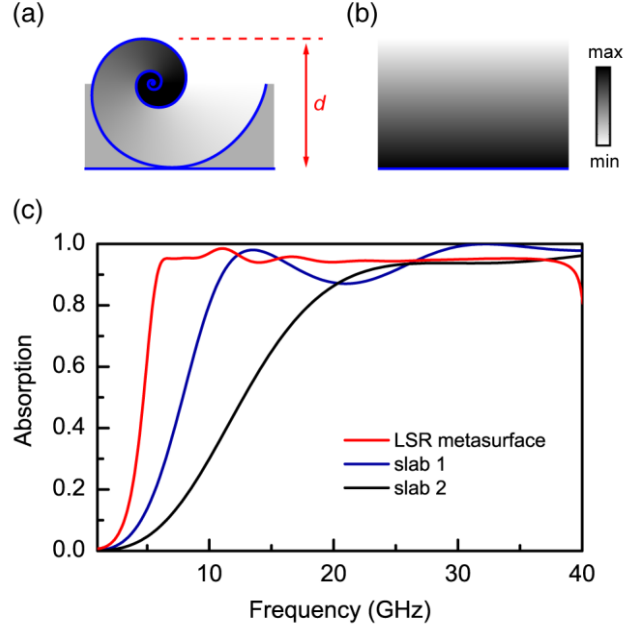

Fig. 3 (a) Unit cell of the optimized LSR metasurface. The first layer of the LSR is filled with a graded-index dielectric material with $\varepsilon_r = 1 + \theta(\varepsilon_c - 1)/(2\pi)$, where $\varepsilon_c = 4.2 + 2.6i$. The inner layers have $\varepsilon_r = \varepsilon_c$. The dielectric outside the LSR has $\varepsilon_r = 2 + 1.5i$. We set $\alpha = 0.008$ mm, $\beta = 0.2$, and $\theta \in [0, 31.4]$. The unit cell has a period of $a = 7.45$ mm and a thickness of $d = 5$ mm. The color indicates the magnitude of the permittivity (b) A graded-index dielectric slab with $\varepsilon_r = 1 + h(\varepsilon_c - 1)/d$, where $h$ is the distance from the upper surface of the slab. (c) Absorption of the optimized LSR metasurface (red), the graded-index dielectric slab with $\varepsilon_c = 4.2 + 2.6i$ (blue), and the graded-index dielectric slab with $\varepsilon_c = 1.4 + i$.

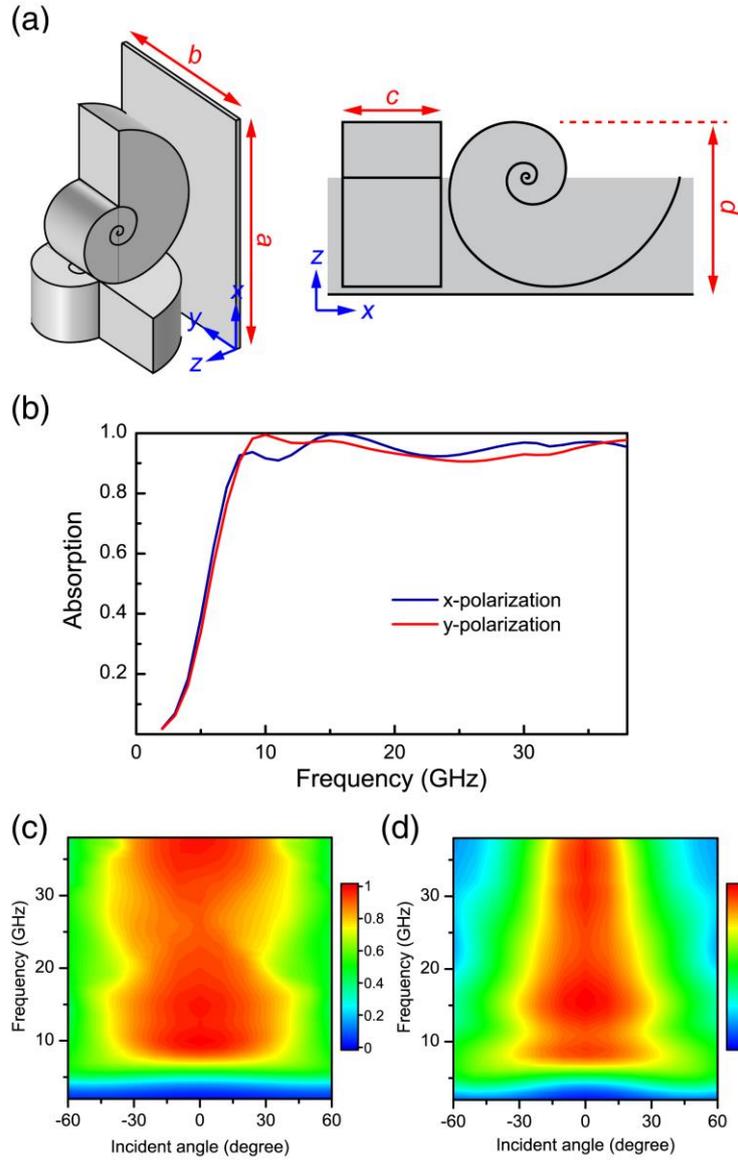

Fig. 4 (a) Unit cell of the LSR metasurface that can absorb both TM- and TE-polarized microwaves. The two LSRs are orthogonal and backed with a metal surface. We set $a$ = 10.2 mm, $b$ = 7.2 mm, $c$ = 2.5 mm, and $d$ = 5.4 mm. (b) Absorption for incident wave polarized along $x$ and $y$ directions. (c) Angle dependence of the absorption for $y$ (i.e. TM) polarization. (d) Angle dependence of the absorption for $x$ (i.e. TE) polarization.

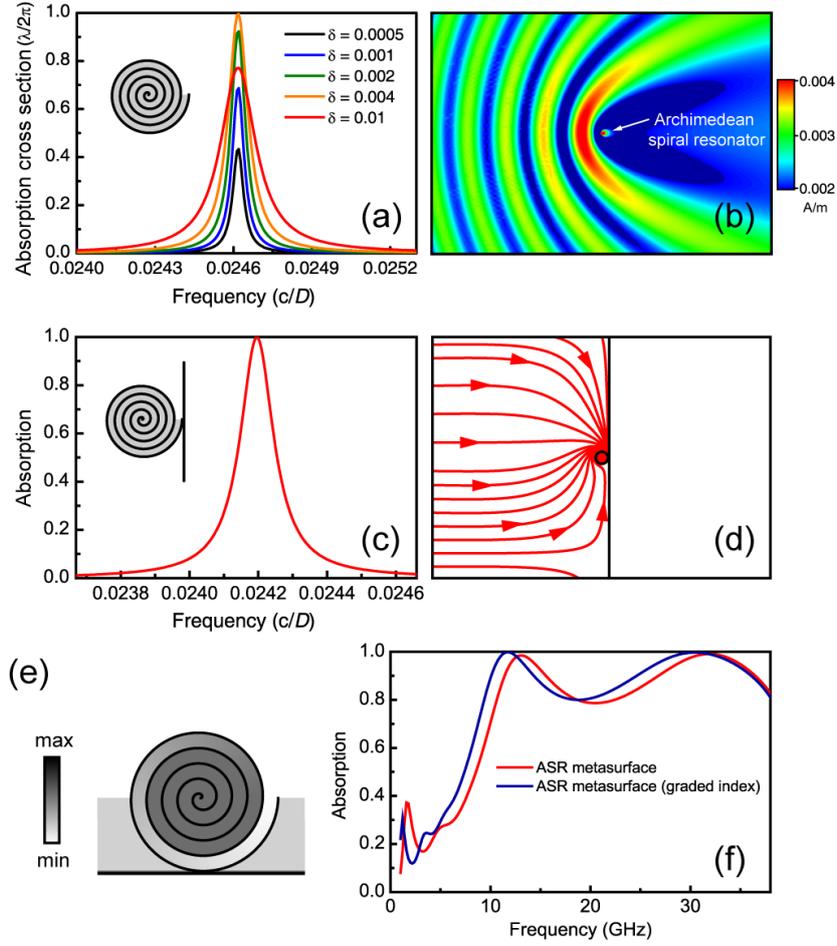

Fig. 5 (a) Absorption cross section of single ASR for dielectric materials with different loss parameters ($\varepsilon_r = 2 + i\delta$). The inset is a schematic of the ASR. (b) Magnetic field amplitude at the resonance frequency. The arrow marks the position of the ASR. (c) Absorption of the metasurface formed of ASRs backed with a metal surface. The metasurface has a period of 18 $D$ with $D$ being the diameter of the ASR. (d) Poynting vectors in one unit cell of the metasurface, showing the singular behaviour of the ASR. (e) Unit cell of optimized ASR metasurface. The first layer is filled with a graded-index dielectric material with $\varepsilon_r = 1 + \theta(\varepsilon_c - 1)/(2\pi)$, where $\varepsilon_c = 4.2 + 2.6i$. The inner layers have $\varepsilon_r = \varepsilon_c$. The dielectric outside the LSR has $\varepsilon_r = 2 + 1.5i$. The thickness and period are the same as the LSR metasurface in Fig. 3(a). (f) Absorption of the ASR metasurface with homogeneous dielectric of $\varepsilon_r = 1.4 + i$ (red) and graded-index dielectric (blue).